\documentclass[aps,preprint,superscriptaddress,showpacs]{revtex4}
\usepackage{epsfig}
\usepackage{natbib}
\usepackage{amsmath}
\usepackage{times}
\usepackage{color}
\usepackage{psfrag}
\usepackage{subfigure}
\newcommand{\beq}{\begin{equation}}
\newcommand{\eeq}{\end{equation}}

\begin{document}

\title{Escape Time Characterization of Pendular Fabry-Perot}

\author{P. Addesso}
\affiliation{Dept. of Electronic and Computer Engineering, University of Salerno, Via Ponte Don Melillo, 1, I-84084 Fisciano, Italy}

\author{V. Pierro}
\affiliation{Dept. of Engineering, University of Sannio, Corso Garibaldi, 107, I-82100 Benevento, Italy}

\author{G. Filatrella}
 \affiliation{Dept. of Sciences for Biological, Geological, and
Environmental Studies\\
\small and Salerno unit of CNSIM, University of Sannio, Via Port'Arsa 11,
I-82100 Benevento, Italy}

\begin{abstract}
We show that an escape from the potential minimum of Fabry-Perot interferometers
can be detected measuring the associated sudden change of reflectivity. We demonstrate that the loss of information that occurs retaining only the sequence of escapes , rather than the full trajectory, can be very mild and can lead to an effective method to reveal the noise intensity or the presence of a coherent signal.
\end{abstract}
\pacs{5.10.Gg,05.10.Ln,07.60.L}
\maketitle

\section{Introduction }
\noindent
Many noisy systems, from condensed matter to nuclear physics, are characterized by a 
sudden passage from an initial state to the final product, the fluctuation-induced escape problem treated by Kramer in a celebrated paper \cite{Kramer40}. 
In such cases one measures the escape times (ETs) to retrieve information about the physical systems, i.e. the activation energy. To restrict the available data to escape events 
rather than to consider the full trajectory might be a simplification of the measurement, but it entails a loss of information, in as much as a continuous variable is substituted by a discrete sequence of events. While in many systems the full trajectory is practically unavailable, i.e. in chemical reactions, in some cases one should balance the advantages of the ETs measurements with the loss of information. 
This is the case, as we will show, for a  Fabry-Perot (FP) cavity. 
We demonstrate how it is possible to simplify the FP measurement detecting the change in reflectivity, at the cost to only retrieve the ETs from the metastable minimum. We also show that the loss of information can be lessened by exploiting the whole ETs distribution with a likelihood ratio test. 

The letter is organized as follows: first we illustrate the possibility to measure the ETs from 
a minimum of the FP cavity. 
Then we explicate the underlying mathematical model that describes the escape process and that 
will be employed to generate the ETs distributions to derive 
the properties of the estimates for the noise intensity and the presence of a coherent sinusoidal are 
outlined. In the final section we conclude and discuss.

We demonstrate how it is possible to simplify the FP measurement detecting the change in reflectivity, at the cost to only retrieve the ETs from the metastable minimum. We also show that the loss of information can be lessened by exploiting the whole ETs distribution with a likelihood ratio test. 

The letter is organized as follows: first we illustrate the possibility to measure the ETs from 
a minimum of the FP cavity. 
Then we explicate the underlying mathematical model that describes the escape process and that 
will be employed to generate the ETs distributions to derive 
the properties of the estimates for the noise intensity and the presence of a coherent sinusoidal are 
outlined. In the final section we conclude and discuss.

\section{Fabry-Perot cavity oscillations: problem statement}
\begin{figure}[tbp]
\centering
\begin{minipage}{14cm}
\begin{picture}(140,140)
\put(0,0.0){\psfig{file=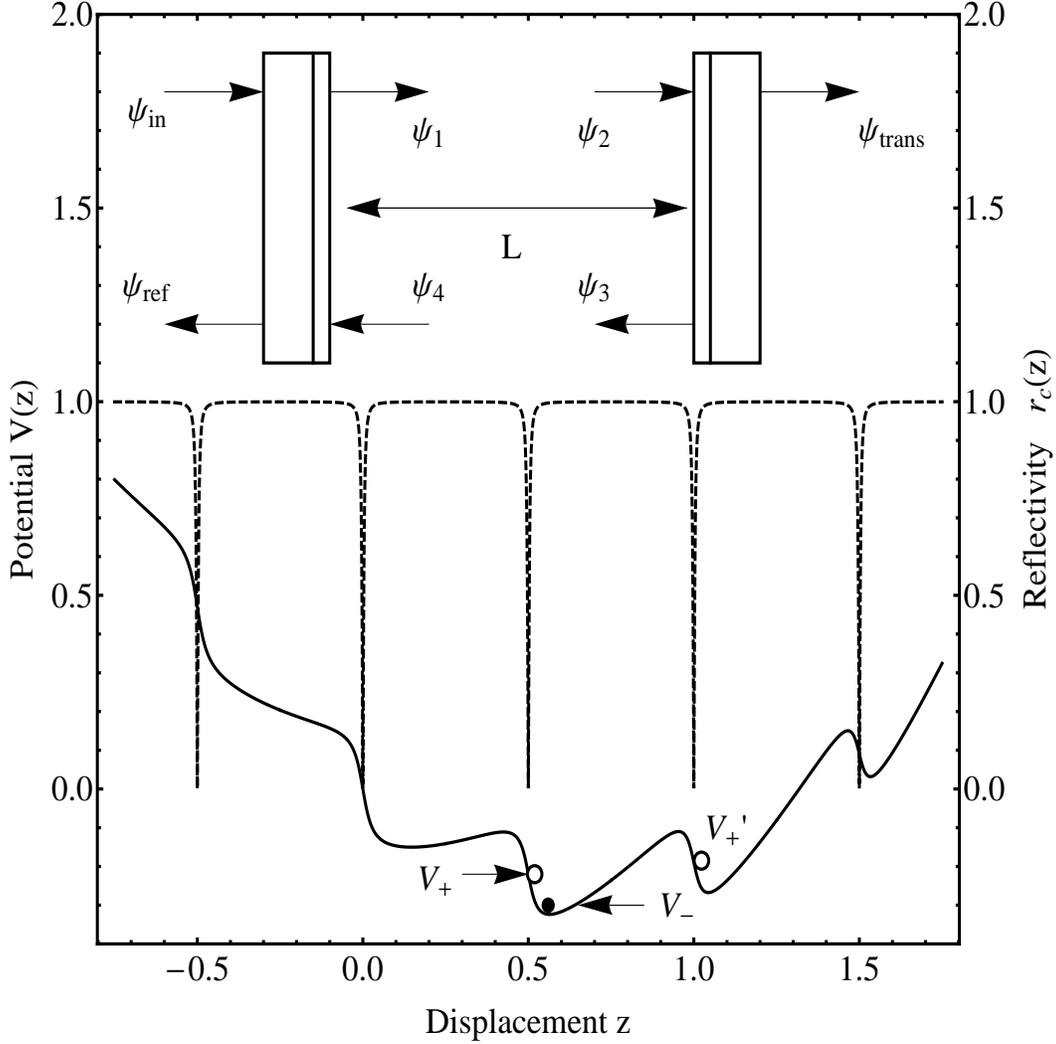
,width=14cm,height=14cm,angle=0.0
}}
\end{picture}
\caption{
The upper part shows the schematic of the mirrors. The incident electromagnetic field $\psi_{in}$ is partially 
reflected ($\psi_{ref}$) and partially transmitted 
The fields at the mirrors consequently read 
$\psi_1=t_1\psi_{in}-r_1\psi_4$, $\psi_{trans}=t_2\psi_1e^{-i\phi}$, $\psi_3=-\psi_2$, $\psi_{ref}=r_1\psi_{in}+t_1\psi_4$.
In the lower part is shown the potential of the { pendular FP } interferometer (solid line) as a function of the displacement 
between the mirrors. The dashed line is the reflectivity, to be monitored to measure ETs. 
The minimum of the potential 
$V_-$ is marked by a black dot, while the minima of the reflectivity (indicating the escape) are 
marked by white dots at the potential $V_+$ and $V_+'$. 
The parameters of the system are: ${\cal F}=50$, $\Pi_{M}=5$, $t_1=t_2=0.97$, ${\cal L}=10^{-6}$.
${\cal L}$ is the loss factor.
}
\label{fig:potential}
\end{minipage}
\end{figure}

A FP cavity is formed by suspended mirrors that make up an optical  cavity for a high power laser. 
The pendulum provides a mechanical restoring force, while the spatial period of the laser radiation pressure induces multistability 
\cite{Deruelle84,Aguirregabiria87,Chickarmane98}. The FP constitutes the latest technology 
in precision measurements of weak signals \cite{Villar10}
 also in the quantum regime \cite{Chan11} or gravitational waves detection  \cite{Rakhmanov98}. 
A weak perturbation forces oscillations of the mirrors that can be detected by the change in the laser power 
accumulated in the optical cavity. 
However, noise (included the noise that enters the system through the measurement devices) should be constrained to 
reveal the very weak interaction of gravitational waves with matter, for noise can induce spurious oscillations of the
 pendulum that mask the effect of the gravitational waves. 
The key feature that we want to exploit is the fact that when the displacement of the pendulums matches a semi-integer
number of the laser light wavelength, a sudden change of reflectivity occurs, see the upper part of Fig. \ref{fig:potential}. It is thus possible to measure the passage across a point close to the maximum of the potential, denoted by $V_+$ and $V_+'$ in the lower part of Fig. \ref{fig:potential}, and 
{to record the time spent in the potential minimum. 
The role of the noise is, in this setup, twofold (as in stochastic resonance \cite{Gammaitoni98}): on one hand it 
helps to overcome the potential barrier and on the other hand the noise covers the signal. 
We propose to characterize the FP through the analysis of the time to escape from a minimum of the optomechanical potential \cite{Marquardt06} to exploit the statistics of occasional large excursions (made possible by the noise) that can be detected as escape events.} 
The advantage of using Escape Times (ETs) stems from  the simplicity of the measurement, because the change of reflectivity can be detected with the accuracy of optical measurements. It is thus possible to record the time that has elapsed between two changes, or to measure an ET from a minimum, with an analysis of the light transmitted through the mirrors without mechanical interaction with the pendulums. 
In other words, the measurement of the reflectance allows us to analyze the dynamics of the pendulums without the application of the servo system, i.e. to follow the almost unconstrained motion. This is a simplification of the measurement technique that could also pave the way for an analysis of the quantum regime \cite{Ludwig08}. 
The simplification has a cost: the ETs abridge the trajectory and therefore some information is lost. However, as has already been pointed out in the case of superconducting Josephson detectors, the loss is relatively mild and can be partially compensated by a careful analysis of the ETs \cite{Addesso12}.

In fact ETs have proved to be effective in evidentiating subtle effects such as the quantum tunneling  
from different attractors \cite{Devoret85},  
the granular nature of the charge \cite{Pekola} and the discrimination between classical and quantum 
activation \cite{Rotoli07} in Josephson junctions, or the critical exponents of bifurcations in micromechanical oscillators 
\cite{Chan07}.

Finally, let us remark that if the system is reset after the crossing of the reflectivity minimum, the measured time corresponds to a first passage time, while it amounts to an escape if the pendulums are left to freely oscillate, for the inclusion of an additional backward probability current \cite{Risken}. However, when the absorbing barrier is sufficiently close to the maximum of the potential the ET and the first passage time are comparable, and there exist formula for the 
passage from one quantity to the other \cite{Reimann99,Boilley04}. 
In the following we use the escape time and first passage time interchangeably, neglecting the difference.

\section{Model equations}

To make these ideas quantitative for FP cavities, let us start with the analysis of the steady state fields 
at the mirror points (see the upper part of Fig. 
\ref{fig:potential}). The FP cavity stores a circulating power $P_{circ}$:
\beq
P_{circ}=|\psi_1|^2 = \chi^2P_{in} {\cal A}(\phi).
\label{eq:circpower}
\eeq
\noindent Here $\chi=P_{M}/P_{in}=t_1^2/(1-r_1 r_2)^2$ denotes the FP gain ($r_i$ and $t_i$ are the reflection and transmission coefficients, $P_{M}$ and $P_{in}$ 
are the maximum and input  laser power, ${\cal A}$ is the Airy function ${\cal A} =1/(1+{\cal F} \sin^2{\phi})$, ${\cal F}=\pi\sqrt{r_1 r_2}/(1-r_1r_2)$ the Finesse of 
the cavity, and  $\phi$ the phase of the light that also determines the half-width $\delta \phi$ of the resonance, $\delta \phi = \pi/2 {\cal F}$ .
The maximum stored power corresponds to the peaks of the Airy function ($\phi=n\pi$). 
The detuning phase can be changed to match the resonance condition by varying either the length of the cavity $L$ or the frequency of the input light. 
We introduce a further simplification by assuming that the mirrors move slow enough that the laser transients can be neglected \cite{Marquardt06}. 

This corresponds to the adiabatic approximation where the electromagnetic transients can be neglected, 
for the distance is large compared to the characteristic time: $2\pi/\omega_0>>(2L/c)|\ln(r_1r_2)|^{-1}$ 
($c$ is the speed of light and $\omega_0$ is the resonant angular velocity of the pendulum).
In the adiabatic limit the laser builds up the energy in the cavity much faster than the mirrors displacements, 
and the energy stored in the cavity is a  function of the positions of the mirrors and not of the previous history. 
Consequently,  the fields inside the cavity are only determined by the positions $x_1$ and $x_2$ 
of the mirrors through the relative coordinate $\tilde{z}=x_2-x_1$ and the phase of the light 
$\phi=2\pi (L + \tilde{z})/\lambda$ ($\lambda$ is the wavelength of the light). The normalized displacement $z=\tilde{z}/\lambda$  
is governed by the following normalized stochastic differential equation 
(for details see  \cite{Pierro94}) :
\beq
\ddot z + \gamma \dot z = - z + \Pi_{M}{\cal A}(2\pi z) + \varepsilon \sin(\Omega t + \phi_0) + \xi(t),
\label{eq:zeta_norm}
\eeq
\noindent where overdot denotes differentiation with respect to the time normalized to $\omega_0 ^{-1}$,
$\Pi_{M} = R_{M}/\mu \lambda \omega_0^2$ is related to the maximum radiation pressure $R_{M} = (2/c)P_{M}$ and $\mu$ is the reduced mass of the mirrors. 
The friction constant $\gamma=\tilde{\gamma}/\omega_0$ is given by the pendulum dissipation constant $\tilde{\gamma}$ divided by
 $\omega_0$. 
The parameter that tunes the nonlinearity is the Finesse, that for interferometers typically lies in the interval  ${\cal F} \simeq 1 \div 1000$.
In normalized units the pendulums are subject to the potential shown in the lower part of Fig. \ref{fig:potential}:
\beq
V(z)=\frac{z^2}{2}- 
\Pi_{M}\frac{\tan^{-1}[ \sqrt{{\cal F}+1}\tan(2\pi z)]+\pi\lfloor 2z+1/2\rfloor }{2\pi\sqrt{{\cal F}+1}}.
\label{eq:pot_norm}
\eeq
In the model Eq. (\ref{eq:zeta_norm}) a sinusoidal forcing term of normalized amplitude $\varepsilon$ and a 
stochastic noise term $\xi$ are included.
In the upper part of Fig. \ref{fig:potential} is also displayed an easily measurable quantity 
introduced and discussed in the previous section:
the FP reflectivity (see \cite{Rakhmanov98} for details).
The reflectivity is a  function of the normalized mirrors displacement through the phase of the light $\phi$:  
\beq
r_c(z):=\left| \frac{\psi_{ref}}{\psi_{in}}\right|=\left|
r_1- \frac{t_1^2 r_2 e^{-2 i\phi}}{1-r_1 r_2e^{-2 i\phi}}
\right|.
\label{eq:reflectVP}
\eeq

The dynamical change of the mirror displacement, { determined by the effect of both noise and the signal, 
}
see Eq.(\ref{eq:zeta_norm}), affects the
reflectivity that abruptly changes in the points indicated in Fig. 
\ref{fig:potential}, as discussed in the previous section.

Let us finally discuss an approximation regarding the noise term $\xi(t)$ introduced in (\ref{eq:zeta_norm}).
In actual  { pendular FP } interferometers the noise is colored, being dominated at low frequency by the seismic noise. The frequency spectrum is affected by several other sources of noise
such as substrate Brownian noise, thermal fluctuations of the expansion coefficient of the test mass, and the noise arising from the mechanical dissipation of the mirrors coating \cite{Bodiya12}. However, to maintain the discussion at a general level, we assume the total external noise term $\xi(t)$ Gaussian and white: $<\xi(t)\xi(t')> = 2 D \delta(t-t')$. 
In this setup, in the next sections we focus on two tasks: the estimation of noise 
intensity $D$ and the detection of a sinusoidal signal.

\section{Noise intensity estimation}
\begin{figure}[tbp]
\centering
\begin{minipage}{12cm}
\begin{picture}(120,120)
\put(0,0.0){\psfig{file=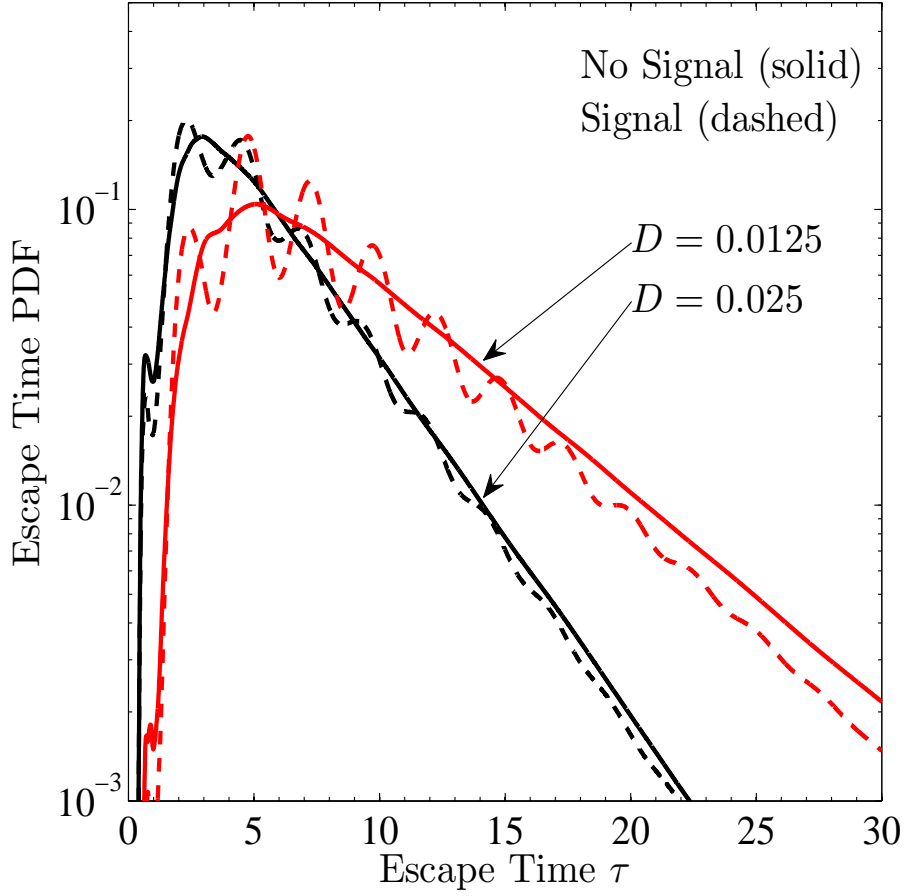,width=12cm,height=12cm,angle=0.0}}
\end{picture}
\caption{
Probability distribution of the ET with (dashed line, $\varepsilon \neq 0$) or without (solid line, $\varepsilon =0$) an applied signal for two different values of the 
noise intensity ($D=0.025$ and $D=0.0125$). The signal amplitude reads $\varepsilon = 0.05$, the phase $\Phi_0 = 0$, and the frequency $\Omega = 2.6$. The parameters of the system are: ${\cal  F} = 2$, $\Pi_{M} = 2.1$, $\Delta V = 0.07$, $\omega_r = 2.9$, $\gamma = 10^{-6}$.
}
\label{fig:PDF}
\end{minipage}
\end{figure}

In this section we propose an estimation procedure for noise level based on escape time statistics.
We note that in the absence of an applied signal ($\varepsilon=0$) the system leaves the potential minima of Eq.(\ref{eq:pot_norm}) 
at the Arrhenius rate  $ \propto \exp{(-\gamma \Delta V / D)}$, for the external noise does not follow the fluctuation 
dissipation theorem \cite{Hanggi82} ($\Delta V=V_+ - V_-$, where $V_+$ and $V_-$ denote the exit point and the minimum 
of the potential, respectively, as shown in the lower part of Fig. \ref{fig:potential}).
The pendular interferometer is characterized by a very low damping term $\gamma \simeq 10^{-6}$ \cite{Drever83}, 
thus the system equation (\ref{eq:zeta_norm})
should be handled in the extremely low dissipation limit. Some analytical approaches have been proposed to extend the 
vanishingly damping treatment of 
stochastic differential equations up to finite values \cite{Landauer83,Drozdov99}. We have employed the method described by Ref. \cite{Melnikov86} to find the ET of Eq.(\ref{eq:zeta_norm}) in
 the absence of the signal, inserting the potential Eq.(\ref{eq:pot_norm}) in the energy diffusion limit of the escape 
 rate \cite{Mazo10}. Using the derivative of the 
action $I$ in the minimum of potential and  integrating by parts we obtain:
\begin{eqnarray}
<\tau> &=& \frac{\gamma}{D^2}\int_{V_-}^{V_+} { I(E)e^{-E\gamma/D} dE \int_{E}^{V_+}
{\frac{e^{-E'\gamma/D}}{I(E')}  dE' }  } + \nonumber \\
&  +& \frac{\Delta V}{D}
\label{eq:integral}
\end{eqnarray}
\noindent that is more convenient for both numerical and analytical treatment. In fact, exploiting the approximation :
\beq
I(E) \simeq (V-V_-)2\pi/\omega_r+ (V-V_-)^2\ddot{I}(V_-)/2,
\label{eq:azione}
\eeq
where $\omega_r$ is the well resonance, the average ET, Eq.(\ref{eq:integral}) can be analytically evaluated :
\beq
<\tau> =(1 +2 \rho) <\tau _H> -\frac{1}{\gamma} (1-2 \rho ) \log (1+\rho  \gamma \Delta V / D) +
\nonumber
\eeq
\beq
+~~\frac{1}{\gamma} e^{-1/\rho } (1+2 \rho) \left[\text{Ei}\left(\frac{1}{\rho }\right)-\text{Ei}\left(\gamma 
\Delta V / D+\frac{1}{\rho }\right)\right].
\label{eq:tau_theoretical}
\eeq
Above, $\text{Ei}(\cdot)$ is the exponential integral function (see \cite{Prudnikov98} for definition and details).
\noindent For $\rho=\omega_r \ddot{I}(V_-) D/(2\pi\gamma) = 0 $, Eq.(\ref{eq:tau_theoretical}) reduces to the ET of the harmonic oscillator :
\beq
<\tau_H>=(\text{Ei}(\gamma \Delta V / D)-\log (\gamma \Delta V / D)-\gamma_E)/\gamma,
\eeq
where $\gamma_E$ is the Euler-Mascheroni constant.
From the numerical distributions of the ETs  shown in Fig.\ref{fig:PDF} it is evident that  ETs distribution can be exploited 
to determine the intensity of the noise. One can compare the average ET with the predictions of Eq.(\ref{eq:tau_theoretical}) 
 to estimate the best fit temperature. This amounts to inferring the noise level from the distributions of Fig. \ref{fig:PDF} with statistical estimation of the parameter $D$ in Eq.(\ref{eq:tau_theoretical}). To evaluate the performances of such procedure we have numerically retrieved the ETs distributions 
 employing a quasi-symplectic modified velocity Verlet algorithm with velocity randomization for the integration of stochastic differential equations 
\cite{Sivak12}. Moreover, we have  found consistent results  with another leapfrog algorithm \cite{Mannella04} that has proved to be very efficient even 
at extremely low dissipation \cite{Burrage07}. Finally, the algorithms have been tested against the known estimates for the washboard potential at very low dissipation ($\gamma << 1$) and noise ($D<<1$) \cite{Melnikov86}. We have found, for the whole dissipation range $10^{-6}<\gamma<10^{-2}$, that the theory lies within the $95\%$ confidence limit of the numerical simulations and  that the distortion of the temperature estimate is below $5\%$. 

\begin{figure}[tbp]
\centering
\begin{minipage}{12cm}
\begin{picture}(120,120)
\put(0,0.0){\psfig{file=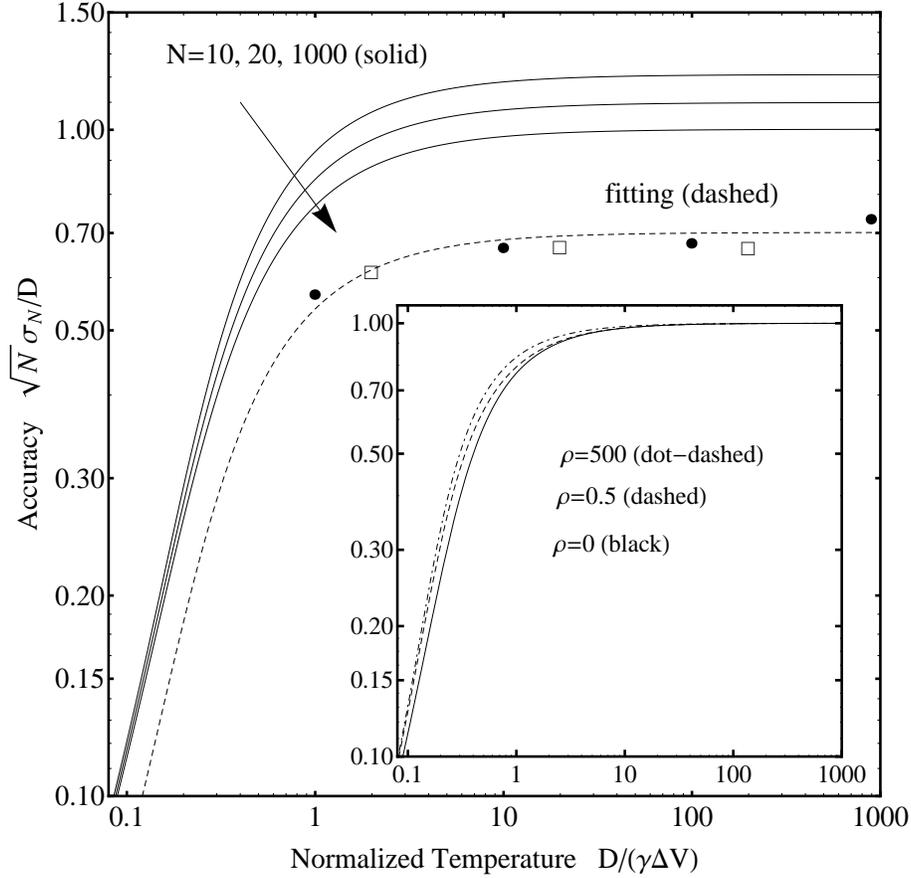,width=12cm,height=12cm,angle=0.0}}
\end{picture}
\caption[] {\footnotesize 
Relative accuracy of the detected noise level as a function of the external noise intensity for different values of the sample numerosity. 
Symbols refer to simulations of the Langevin Eq.(\ref{eq:zeta_norm}) for $N=1000$, ${\cal F} = 20$, $\Pi_M = 2$, $\Delta V = 0.042$, $D=0.00042$ (filled points) and $D=0.00084$ (empty circles). The dashed line denotes the fitting. The inset shows the asymptotic
 behaviors as a function of the parameter $\rho$ that represents the deviations of the model Eq.(\ref{eq:zeta_norm}) from the
 harmonic oscillator ($\rho=0$).
 }
\label{fig:noisedetection}
\end{minipage}
\end{figure}

The efficacy of the method for the estimate of the noise level has been evaluated computing the estimator variance $\sigma^2_N$ as a function of the sample size $N$ and 
of the noise intensity $D$, under the hypothesis that ETs are described by an exponential distribution.
This estimator is the Maximum Likelihood estimator, and its large $N$ behavior can be determined according to 
\begin{equation}
\sigma_N^2 \simeq \frac{\left[\mathcal{D}'(<\tau>)\right]^2}{NJ(<\tau>)},
\end{equation}
where the function $\mathcal{D}(<\tau>)$ is the inverse of Eq.(\ref{eq:tau_theoretical}) w.r.t. the noise level $D$, 
and  $J(<\tau>)$ is the Fisher information , that for the exponential distribution reduces to
 $J(<\tau>) = 1/(<\tau>)^2$ \cite{Lehmann99}.
In Fig. \ref{fig:noisedetection} we show that the relative error 
increases with noise level up to a saturation point (around $D\simeq \gamma \Delta V$) that depends, as expected, upon $N^{-1/2}$ and exhibits a weak dependence
 upon $\rho$ (see the inset).
The estimate of the noise level improves by increasing $\gamma\Delta V$ i.e. 
the dissipation and/or the potential barrier. From the physical consideration that a minimum time occurs before escape \cite{Berglund05} (see also Fig. \ref{fig:PDF}), it is evident that the ET density departs from the exponential model for short escapes. Such deviations from the exponential distribution lead 
to an overestimate of the relative error (as evident from the fitted curve in Fig. \ref{fig:noisedetection}), 
since the numerical result is about $30\%$ smaller.

\section{Sinusoidal signal detection}
We shall now focus on sinusoidal perturbations, i.e. $\varepsilon \neq 0$ in Eq.(\ref{eq:zeta_norm}). A sinusoidal signal  modifies the escape process \cite{Jung93,Berglund05,Filatrella10} (see Fig.\ref{fig:PDF}) and could therefore be revealed with an analysis of the ET distributions. 
The analysis  can be performed with a direct Sample Mean (SM) procedure that considers the average ET. The SM detection of the signal is performed observing a change of the average ($<\tau_\varepsilon>$ for $\varepsilon \neq 0$ instead of $<\tau>$), a method that has been used to investigate the effect of a periodic signal in Josephson junctions \cite{Yu03,Sun07}. 
{ We underline the role of the noise: was it be absent, the system would never experience an escape over the barrier, ($<\tau>
\rightarrow\infty$ in the zero noise limit). Therefore the signal 
induced displacements are much smaller than the light wavelength, 
as required to make the detection method interesting.
}

\begin{figure}[htb]
\centering
\begin{minipage}{10cm}
\begin{picture}(100,70)
\put(0,0.0){\psfig{file=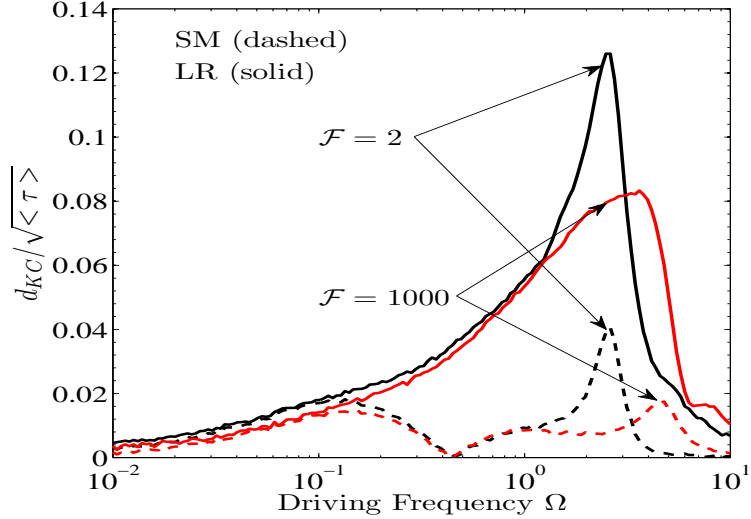,width=10cm,height=7cm,angle=0.0}}
\end{picture}
\caption[] {\footnotesize 
Normalized Kumar-Carroll index as a function of the angular velocity of the applied signal for the SM approach (dashed line)
 and the Likelihood Ratio (solid line) keeping the same barrier height $\Delta V=0.07$
for two Finesse parameters : $~{\cal F}=2$ ($<\tau> = 9.4$, $\Pi_{M}=2.1$ and $\omega_r=2.9$) and ${\cal F}=1000$ ($<\tau> = 8.5$, $\Pi_{M}=12$ and
 $\omega_r=6.6$). 
The parameters of the system are: $\varepsilon= 0.05$, $\Phi_0 = 0$, $D = 0.0125$, $\gamma = 10^{-6}$. As a reference, the normalized $d_{KC}$ 
index for the adapted filter is independent of the frequency and reads  $\varepsilon/(2\sqrt{D})= 0.22$.
}
\label{fig:NDKC}
\end{minipage}
\end{figure}

The SM escape can be optimized with an 
appropriated choice of the detector characteristics that depends upon the noise intensity, and the signal
properties to achieve the matching between the noise induced ET and the signal period 
 (see the lower part of Fig. \ref{fig:NDKC}), around $\Omega \simeq 10^{-1}$) \cite{Yu03,Filatrella10}:
\beq
<\tau>\simeq 2\pi/\Omega.
\label{eq:matching}
\eeq

\noindent Eq. (\ref{eq:matching}) ensures a resonant activation, i.e. the possibility to exploit noise to facilitate signal detection \cite{Gammaitoni98}. 
To improve the detection, we have employed the Likelihood Ratio (LR) technique that analyzes the data exploiting the full distribution of the ETs. 
 LR amounts to performing a statistical test to distinguish if the finite sample collected by the experiments is better fitted by the unperturbed escape ($\varepsilon = 0$) or by an oscillating distribution ($\varepsilon \neq 0$). It consists in averaging the samples of a random variable $\Lambda$ obtained from escapes by the nonlinear mapping
\begin{equation}
\Lambda = \ln\left[\frac{f_{\varepsilon}(\tau)}{f(\tau)} \right],
\end{equation}
where the probability density functions $f_{\varepsilon}(\tau)$ and $f(\tau)$ of the escapes are numerically computed with an approach  similar to the analysis of the ETs of Josephson junctions \cite{Addesso12}.
The LR leads to a remarkable improvement of the performances with respect to the analysis of the average ET, 
because the averages $<\Lambda>$ and $<\Lambda_{\varepsilon}>$ (computed for $\varepsilon = 0$ and $\varepsilon \neq 0$ respectively) carry much more information.
This is shown in Fig.\ref{fig:NDKC} through the index $d_{KC}$ \cite{Kumar84}, a rough measure of the efficiency of the detection method that, for $Y = \{\tau, \Lambda \}$ and $Y_{\varepsilon} = \{\tau_{\varepsilon}, \Lambda_{\varepsilon} \}$ (related to SM and LR respectively), 
reduces to
\begin{equation}
d_{KC}=\frac{|<Y_{\varepsilon}>-<Y>|}{\sqrt{\frac{1}{2}\left[\sigma^2(Y_{\varepsilon})+\sigma^2(Y)\right]}},
\label{eq:dKC}
\end{equation}
where $\sigma(Y_{\varepsilon})^2$ and $\sigma(Y)^2$ refer to the variances of $Y_{\varepsilon}$ and $Y$ respectively
(more details are in \cite{Addesso12}).
Moreover we take into account the time needed to observe an escape by normalizing the index $d_{KC}$ w.r.t. 
the square root of the average ET (computed in the case of $\varepsilon = 0$). 
Thus we have a rough measure of the amount of information per unit time extracted by the detection strategy.
As clearly shown in Fig. \ref{fig:NDKC}, in the most favorable condition, i.e. a signal frequency close to resonant 
frequency of the potential well, the index improvement for LR is as much as four-times that of the SM.
The improvements are over the whole spectrum of the external signal, and in particular the performances increase at
a frequency above the matching Eq.(\ref{eq:matching}), while the two methods have the same efficiency below 
$\Omega=0.2$. With the LR the stochastic resonant peak of Eq.(\ref{eq:matching}) disappears as expected \cite{Addesso12}. The loss of information in the best frequency condition (for a low dissipation FP cavity) with respect to the matched filter is roughly a factor $2$ at low Finesse values (${\cal F} = 2$) and increases to a factor $3$ at high values (${\cal F}=1000$). 
Putting it another way, the analysis of ETs requires about $4 \div 9$ times more data than the optimal matched filter in order to
obtain the same detection performances.
The number of escapes is an essential aspect of the  
detection of small signals ($\varepsilon << \Delta V$). When the signal is weak the number increases and thus constitutes a severe 
limitation to the detection. We have numerically checked that, at a reasonably high level of the index (\ref{eq:dKC}), $d_{KC}=8$, the number of escapes increases as the square of the ratio between the noise intensity and the signal amplitude $\varepsilon$, as for the washboard potential \cite{Addesso12}. For instance to decrease the signal amplitude from $\varepsilon=0.05$ (as in Fig.\ref{fig:NDKC}) to $\varepsilon = 0.0005$ requires to increase the number of escapes from $N\simeq 400$ to $N\simeq 3\times 10^6$ for the lower value of the Finesse (${\cal F}=2$), and from $N\simeq 900$ to $N\simeq 7\times 10^6$ for the higher value of the Finesse (${\cal F}=1000$).


\section{Conclusions and Outlooks}
We have shown that escape times can be effectively employed to determine the noise intensity and the presence of a 
coherent signal affecting a metastable system. To do so, we have focused on a FP interferometer, 
and we have explicitly discussed the advantages of the detection of the escape times: being mechanically decoupled, 
such characterization is in fact promising for measurements in the semiclassical or quantum regime. 
In spite of these advantages, had the escape sequence entailed a great loss of information respect to the full 
trajectory, the ET reduction would not be appealing. 
For this reason we have accurately compared the performances of the ET analysis with the full trajectory analysis 
(analysis performed through a matched filter, that can be demonstrated to be the optimal method). 
We have found that: 1) The noise intensity estimate is not distorted (within $5\%$); 
2) The noise estimate exhibits the optimal behavior $N^{1/2}$ for large $N$; 3) The estimates are more
accurate in the limit of rare ETs;  
4) The detection of a sinusoidal signal with the analysis of the 
average escape rate entails a loss of information (measured by the index $d_{KC}$) of about an order of magnitude; 
5) The detection of a coherent signal with a more accurate analysis based on a likelihood ratio estimator reduces the 
loss of the index $d_{KC}$ to a factor $2 \div 3$; 
6) Signal enhancement at a matching finite temperature given by Eq.(\ref{eq:matching}), or stochastic resonance, is 
apparent in the analysis of the average escape time only. The more refined analysis with likelihood estimator enhances 
the performances that monotonically improve moving towards the resonant frequency, see Fig.\ref{fig:NDKC}; 
{ 7) The detection sinusoidal signals exhibits the optimal behavior $N^{1/2}$ for large $N$, however the detection of small signals can be very challenging for mechanical systems, because the system is inherently slow and it might require an unpractical number of periods -- as opposite to electronic systems such as Josephson junctions that are much faster and therefore potentially more suitable \cite{Filatrella10}. }

These findings are of interest {\it per se} for the applications of Fabry-Perot pendular cavities in  {fundamental } physics. However, a natural question that arises is about their generality for other systems that can be characterized by ETs. In other words, one could ask the following questions: does the loss of information (as measured by the variance of the estimate) entailed in the usage of ET depends upon the particular shape of the metastable potential? We lack a general answer, however one could notice that these features are also shared by the washboard potential associated to Josephson junctions \cite{Addesso12}, and therefore we conjecture that they might be a general feature of estimates obtained through ETs. 

Moreover, there are two findings that are specific for ETs from FP potential, eq.(\ref{eq:pot_norm}). First, the analytic evaluation of the average ET, eq.(\ref{eq:tau_theoretical}), together with the assumption of an exponential distribution, captures the main features of the noise estimate with ETs, although with a discrepancy (around $30\%$) of the standard deviation, see Fig. \ref{fig:noisedetection}. This finding is therefore useful to evaluate the performances of the estimates with ETs as a function of the FP parameter without heavy stochastic numerical simulations. Second, the nonlinearity parameter (the Finesse $\cal F$) deteriorates the performances of FP as coherent signal detector, see Fig. \ref{fig:NDKC}, thus giving a qualitative suggestion on FP design. 

We underline that in this paper we only illustrate the principle of operation of the ET detection through FP based interferometers, and we have shown the potentiality of the full analysis of the ET statistics.
Essentially we suggest that the ET measurements method could be interesting if a favorable balance between operational simplicity and information loss can be achieved.

\section{Acknowledgments}
This work has been supported by the Italian Super Computing Resource Allocation 
ISCRA, CINECA, Italy (Grant IscrB\_NDJJBS 2011).


\newpage

\begin{thebibliography}{30}

\bibitem{Kramer40}
H. A. Kramers, Physica (Utrecht) {\bf 7}, 284 (1940).

\bibitem{Deruelle84} 
N.  Deruelle and P. Tourrenc, {\it Gravitation, Geometry and Relativistic Physics}, (Springer-Verlag, Berlin), 1984.

\bibitem{Aguirregabiria87} 
J.M. Aguirregabiria and L. Bel, Phys. Rev. A, {\bf 36}, 3768 (1987).

\bibitem{Chickarmane98} 
V. Chickarmane,  Dhurandhar S. V.,  Barillet R.,  Hello P. and  Vinet J.-Y., 
Appl. Opt. {\bf 7}, 3236 (1998).

\bibitem{Villar10} 
A.E. Villar, E. D. Black  , DeSalvo R., Libbrecht K. G.,Marquardt F., Michel C.,  Morgado N., Pinard L., Pinto I. M., Pierro V., Galdi V., Principe M., and Taurasi I., Phys. Rev. D {\bf 81}, 122001 (2010).

\bibitem{Chan11}
J. Chan, Mayer Alegre T.P., Safavi-Naeini A.H., Hill J.T., 
Krause A., Gr\"oblacher S., Aspelmeyer M. and Painter O.
Nature {\bf 478},  89 (2011).

\bibitem{Rakhmanov98} 
M. Rakhmanov, { \it Dynamics of Fabry-Perot resonators with suspended mirrors. 2. Delay effects and control system} , 
LIGO technical report T970230 California Institute of Technology, 1998.

\bibitem{Gammaitoni98} 
L. Gammaitoni , H\"anggi P., Jung P. and Marchesoni F. Rev. Mod. Phys. {\bf 70},  223 (1998).

\bibitem{Marquardt06}
F. Marquardt,  Harris J. G. E. and Girvin S. M. Phys. Rev. Lett. {\bf 96},  103901 (2006).

\bibitem{Ludwig08} 
M. Ludwig, Kubala B. and Marquardt F., New. J. Phys. {\bf 10},  095013 (2008).

\bibitem{Addesso12} 
P. Addesso, Filatrella G. and Pierro V., Phys. Rev. E {\bf 85}, 016708 (2012).

\bibitem{Devoret85}
M.H. Devoret, Martinis J. M. and Clarke J., Phys. Rev. Lett. {\bf 55}, 1908 (1985).

\bibitem{Pekola} 
J. P. Pekola, Phys. Rev. Lett. {\bf 93}, 206601 (2004).

\bibitem{Rotoli07}
G. Rotoli, Bauch T., Lindstrom  T., Stornaiuolo  D., Tafuri F. and Lombardi F., 
Phys. Rev. B {\bf 75}, 144501 (2007).

\bibitem{Chan07} 
H.B.Chan and Stambaugh C.,
Phys. Rev. Lett. {\bf 99}, 060601 (2007).

\bibitem{Risken}
H. Risken, {\it The Fokker-Planck Equation: Methods of Solution and Applications}, Springer, Berlin, (1989).

\bibitem{Reimann99}
P. Reimann, Schmid G. J. and Hanggi P.,
Phys. Rev. E {\bf 60}, R1 (1999).

\bibitem{Boilley04}
D. Boilley, Jurado B. and Schmitt C.,
Phys. Rev. E {\bf 70},  056129 (2004).

\bibitem{Pierro94} 
V. Pierro and Pinto I. M., Phys. Lett. A {\bf 185}, 14 (1994).

\bibitem{Bodiya12}
T. P. Bodiya, Harry G. and DeSalvo R., {\it Optical coatings and thermal noise in Precision measurements},  
Cambridge Univ. Press, New York, (2012).

\bibitem{Hanggi82} 
P. Hanggi and Thomas H., 
Phys. Rep. {\bf 88},  207 (1982).

\bibitem{Drever83} 
R. Drever, {\it Gravitationa Radiation}, N. Deruelle and T. Piran eds., North-Holland, Amsterdam (1983).

\bibitem{Landauer83} 
M. B\"uttiker, Harris E. P. and Landauer R.,
Phys. Rev. B {\bf 28}, 1268 (1983).

\bibitem{Drozdov99} 
A.N. Drozdov and Hayashi S.,
Phys. Rev. E {\bf 60}, 3804 (1999).

\bibitem{Melnikov86} 
V.I. Melnikov and Meshkov S.,
J. Chem. Phys. {\bf 85}, 1018 (1986) .

\bibitem{Mazo10} 
J. J. Mazo, Naranjo F. and Zueco D.,
Phys. Rev. B {\bf 82}, 094505  (2010).

\bibitem{Prudnikov98}
A.P: Prudnikov,Brychkov Yu. and Marichov O. I., { \it Integrals and Series}, Gordon and Breach, India (1998).

\bibitem{Sivak12} 
D.A. Sivak,Chodera J. D. and Crooks G. E.,
{\it Driven Langevin dynamics: heat, work and pseudo-work} (http://arxiv.org/abs/1107) (2012).

\bibitem{Mannella04} 
R. Mannella, Phys. Rev. E {\bf 69}, 041107  (2004).

\bibitem{Burrage07} 
K. Burrage, Lename I. and Lythe G.,
SIAM J. Sci. Comput. {\bf 29}, 245 (2007).

\bibitem{Lehmann99} 
E.L. Lehmann, {\it Elements of Large-Sample Theory}, Springer-Verlag, New York (1999).

\bibitem{Berglund05} 
N. Berglund and Guentz B.,
Europhys. Lett. {\bf 70}, 1 (2005).

\bibitem{Jung93} 
P. Jung, Phys. Rep. {\bf 234}, 175 (1993).

\bibitem{Filatrella10} 
G. Filatrella and Pierro V.,
Phys. Rev. E {\bf 82}, 046712 (2010).

\bibitem{Yu03} 
Y. Yu and Han S.,
Phys. Rev. Lett. {\bf 91}, 127003 (2003).

\bibitem{Sun07} 
G. Sun, Dong N., Mao G.,Chen J.,Xu W., Ji Z., Kang L.,
Wu P., Yu Y. and Xing D.,
Phys. Rev. E {\bf 75}, 021107 (2007).

\bibitem{Kumar84}
 B. V. K. V. Kumarand Carroll C. W., Opt. Eng.{\bf 23}, 732 (1984).




\end{thebibliography}
\end{document}